\documentclass[conference]{IEEEtran}

\AtBeginDocument{%
  \providecommand\BibTeX{{%
    Bib\TeX}}}

\usepackage{graphicx}
\usepackage{textcomp}
\usepackage{multirow}
\usepackage{xcolor}
\usepackage{pgfplots}
\usepackage{booktabs}
\usepackage[caption=false,font=footnotesize]{subfig}
\pgfplotsset{compat=1.18}
\usepackage{soul}
\usepackage{amsfonts}
\usepackage{array}
\usepackage{setspace}
\usepackage{url}
\usepackage{bbm}
\usepackage{threeparttable}

\usepackage{makecell}

\usepackage{tikz}
\usepgfplotslibrary{statistics,fillbetween}

\usepackage{amsmath}
\usetikzlibrary{arrows.meta, positioning, fit, backgrounds, calc}

\usepackage{siunitx}
\usepackage{tabularx}

\usepackage[ruled, linesnumbered, vlined, longend]{algorithm2e}
\SetAlFnt{\footnotesize}
\SetAlCapNameFnt{\small}
\SetAlCapFnt{\small}
\IncMargin{3pt}
\SetAlgoNlRelativeSize{-0.5}
\SetKwProg{Def}{def}{$\,$:}{}
\SetKwProg{Func}{Func}{$\,$:}{}
\SetKwInOut{Input}{Input}
\SetKwInOut{Output}{Output}

\def\BibTeX{{\rm B\kern-.05em{\sc i\kern-.025em b}\kern-.08em
    T\kern-.1667em\lower.7ex\hbox{E}\kern-.125emX}}

\newcommand{\tighttable}{\setlength{\tabcolsep}{2pt}\renewcommand{\arraystretch}{0.89}}

\newcommand{\ind}{\mathbf{1}}

\usepackage{eso-pic}
\newcommand{\ieeecopyrightnotice}{%
\AddToShipoutPictureBG*{%
\AtPageUpperLeft{%
\put(\LenToUnit{0.5\paperwidth},\LenToUnit{-0.42in}){%
\makebox[0pt][c]{\parbox{\textwidth}{\centering\scriptsize
O.~Cassidy, M.~Andronic and G.~A.~Constantinides, ``ExitLUT: Boolean Early Exits for Efficient FPGA-Based Neural Network Inference,'' in \textit{2026 International Conference on Field Programmable Technology (ICFPT)}, 2026.\\
\copyright~2026 IEEE. Personal use of this material is permitted. Permission from IEEE must be obtained for all other uses, in any current or future media, including reprinting/republishing this material for advertising or promotional purposes, creating new collective works, for resale or redistribution to servers or lists, or reuse of any copyrighted component of this work in other works.}}}}}}

\begin{document}
\ieeecopyrightnotice

\title{ExitLUT: Boolean Early Exits for Efficient FPGA-Based Neural Network Inference}

\author{
Oliver Cassidy, Marta Andronic and George A.~Constantinides\\
Department of Electrical and Electronic Engineering,
Imperial College London, UK\\
Email: \{olly.cassidy23, marta.andronic18, g.constantinides\}@imperial.ac.uk
}

\maketitle

\begin{abstract}

Early exit neural networks reduce average inference cost by terminating once sufficient confidence is reached. On FPGAs, conventional confidence-based exits add arithmetic control paths poorly matched to LUT fabrics, consuming resources that could otherwise support the main datapath. We present ExitLUT, which replaces softmax-style exit decisions with learned Boolean functions mapped directly into FPGA LUTs. In our CNN evaluation, fpgaConvNet is the static spatial-dataflow baseline, while ATHEENA is its softmax-based early exit extension. ExitLUT reduces ATHEENA's exit-decision LUT usage by 2.58$\times$ over the softmax path and eliminates decision-path DSP usage, while preserving accuracy and exit behavior. Rerunning design-space exploration reinvests these savings into the datapath, improving throughput by up to 4.08$\times$ over fpgaConvNet and 1.82$\times$ over ATHEENA. For LUT-based NNs, embedded ExitLUT exits achieve 1.41--1.89$\times$ average time-to-decision speedup over static NeuraLUT-Assemble designs with under 1$\%$ LUT overhead.

\end{abstract}

\section{Introduction}

Deep neural networks (DNNs), including convolutional neural networks (CNNs), are widely used to approximate highly nonlinear functions in tasks such as image classification, detection, and signal processing~\cite{nns,shallowdeep}. Their accuracy is typically achieved by evaluating a fixed sequence of learned transformations, often across many layers. However, conventional static inference executes the same full computation graph for every input, regardless of sample difficulty. This creates avoidable computation: many inputs can be classified correctly using early or intermediate features, yet they are still propagated through all remaining layers, increasing latency, energy consumption, and hardware resource usage~\cite{branchy}.

Early exit mechanisms reduce this inefficiency by attaching auxiliary classifiers to intermediate layers and terminating inference once an intermediate prediction is considered sufficiently reliable. In conventional DNN and CNN accelerators, this decision is usually based on confidence estimates such as maximum softmax probability, entropy, or classification margin, together with manually selected thresholds~\cite{branchy}. While effective in arithmetic-rich software or GPU settings, these methods introduce additional reductions, comparisons, score computations, and control-flow logic. 
On field-programmable gate arrays (FPGAs), this creates an architectural mismatch: a one-bit routing decision is produced by a floating-point confidence pipeline, which can increase latency and resource usage and offset the benefit of early termination~\cite{ATHEENA}.

In this paper, we present ExitLUT, a method for implementing low-overhead early exits in FPGA neural network (NN) accelerators by expressing exit decisions as Boolean functions mapped directly into LUT fabric. By replacing softmax-based confidence estimation with LUT-based decision logic, ExitLUT aligns early exit control with the native computation model of FPGA fabrics.
This removes arithmetic-heavy control from the critical path and frees resources for increased datapath parallelism, while preserving low-latency execution.

ExitLUT demonstrates the advantages of Boolean early exits across two complementary NN backbones deployed on FPGAs. First, for conventional dataflow fixed-point FPGA CNN accelerators, we target ATHEENA, an early exit extension of the fpgaConvNet toolflow. ATHEENA implements branch decisions using a floating-point softmax confidence pipeline. We replace this pipeline with a learned LUT-based exit predictor trained to approximate the original thresholded softmax decision from quantized logits. This reduces exit-decision latency from 180 cycles to 12 cycles and eliminates DSP and BRAM usage. Second, for LUT-based neural networks, where even small external decision circuits can dominate latency, we embed exit conditions directly into the L-LUT representation through auxiliary training and hardware-aware threshold optimization. Each exit emits class flags directly from LUT logic, avoiding softmax and cross-class confidence evaluation. Since the datapath remains spatially instantiated, resource usage is not expected to decrease; the benefit is lower average latency with near-zero area overhead.

Across equivalent design points, replacing ATHEENA's softmax exit path uses 2.58$\times$ fewer exit-decision LUTs than ATHEENA and improves throughput by up to $4.08\times$ over fpgaConvNet and $1.82\times$ over ATHEENA. In LUT-based NNs, relative to the static NeuraLUT-Assemble baseline, ExitLUT speeds up the average latency by 1.41--1.89$\times$ with minimal accuracy loss and under $1\%$ LUT overhead.

In summary, the key contributions of this paper are:

\begin{itemize}
    \item We introduce ExitLUT, an open-source\footnote{\url{https://github.com/ollycassidy13/ExitLUT}} toolflow for implementing early exits using LUT-based Boolean logic.

    \item We instantiate ExitLUT in two FPGA-based backbones:
    \begin{itemize}
        \item For dataflow fixed-point CNN accelerators, we replace ATHEENA's floating-point softmax exit pipeline with a learned LUT-based decision network.

        \item For LUT-based NNs, we train auxiliary exits and fold
        class-specific threshold decisions directly into L-LUT truth tables.
    \end{itemize}

    \item For ATHEENA-style CNN accelerators, ExitLUT improves throughput by
    up to $4.08\times$ over fpgaConvNet and $1.82\times$ over ATHEENA,
    while preserving accuracy, exit probability and high sample-wise agreement.

    \item We show that this gain is enabled by cheaper exit control:
    ExitLUT uses 2.58$\times$ fewer exit-decision LUTs than
    ATHEENA and eliminates decision-path DSP usage.

    \item For LUT-based NNs, our exits with thresholds achieve a 1.41--1.89$\times$ average per-sample latency speedup over
     static NeuraLUT-Assemble models, with
    under $1\%$ LUT-area overhead and minimal accuracy degradation.
\end{itemize}

\section{Background}\label{sec:background}

\subsection{FPGA-based Neural Networks}
\label{subsec:fpga_nn}

CNN-to-FPGA toolflows generally use either time-multiplexed or spatial dataflow architectures. Time-multiplexed designs reuse compute engines across layers, reducing area at the cost of throughput. Spatial dataflow designs instead instantiate dedicated pipeline stages, allowing different layers to process different inputs concurrently. Common compiler-driven examples include fpgaConvNet~\cite{fpgaconvnet}, FINN~\cite{finn}, hls4ml~\cite{hls4ml,hls4ml-bin}, and domain-specific pipelined CNN frameworks such as HPIPE~\cite{hpipe} and H2PIPE~\cite{h2pipe}.

FINN~\cite{finn} and hls4ml~\cite{hls4ml,hls4ml-bin} provide complementary examples of quantization-aware FPGA inference. FINN maps binary and low-bitwidth Quantized Neural Networks (QNNs) onto LUT and DSP resources, while hls4ml generates HLS implementations with configurable numerical precision. 

fpgaConvNet~\cite{fpgaconvnet} is representative of spatial dataflow CNN acceleration and forms the basis of ATHEENA~\cite{ATHEENA}. It represents a trained ONNX model as a Synchronous Dataflow Graph (SDFG), where nodes are parameterized compute blocks and edges are streaming channels. A simulated-annealing Design Space Exploration (DSE) then selects folding factors under resource constraints. Coarse-grain folding time-multiplexes layers, while fine-grain folding adjusts intra-layer parallelism, giving a flexible throughput--area trade-off that must be balanced to avoid bottleneck stages.

Other works such as HPIPE~\cite{hpipe} and H2PIPE~\cite{h2pipe} instead emphasize domain-specific compilation for layer-specialized CNN pipelines, including resource partitioning and data-movement scheduling. Together, these frameworks show the importance of compiler-controlled parallelism, quantization and dataflow scheduling in FPGA NN inference.

\subsection{LUT-based Neural Networks}
\label{subsec:lut_dnn}

In an FPGA, a physical LUT (P-LUT) is a configurable logic primitive that implements an arbitrary Boolean function over a small number of input bits. LUT-based NNs exploit this by mapping neurons through enumeration to logical LUTs (L-LUTs): model-level Boolean functions that are later mapped by logic synthesis onto P-LUTs and routing resources with pipeline registers inserted between layers~\cite{logicnets,poly}.

Several recent works build neural networks directly from LUT-based structures for efficient FPGA inference. LogicNets~\cite{logicnets} learns sparsely connected neuron functions as L-LUTs, making each neuron's truth table tractable for synthesis and implementation. PolyLUT~\cite{poly,assemble2} absorbs polynomial functions into L-LUTs to increase local expressivity, while NeuraLUT~\cite{NeuraLUT} absorbs small subnetworks into each L-LUT. NeuraLUT-Assemble~\cite{assemble} composes trees of these blocks to increase effective fan-in while keeping individual LUT sizes tractable. Related LUT-oriented approaches include decision tree inference in TreeLUT~\cite{treelut}, ensemble-based LUT inference in AmigoLUT~\cite{amigolut}, weightless neural networks~\cite{dwn}, Kolmogorov-Arnold Networks in KANELE~\cite{kanele}, and LUT-aware heterogeneous quantization in HGQ-LUT~\cite{hgqlut}. Prior works~\cite{yukio,compressedlut,reducedlut} also show that post-training modification of L-LUT contents can improve area efficiency.

These works share two properties. First, inference can be extremely low latency because it reduces to combinational LUT evaluation rather than multi-cycle arithmetic. Second, execution is static: every input follows the same hardware path, independent of sample difficulty. Resources are therefore provisioned for the worst case, and the latency--throughput trade-off is fixed at synthesis time.

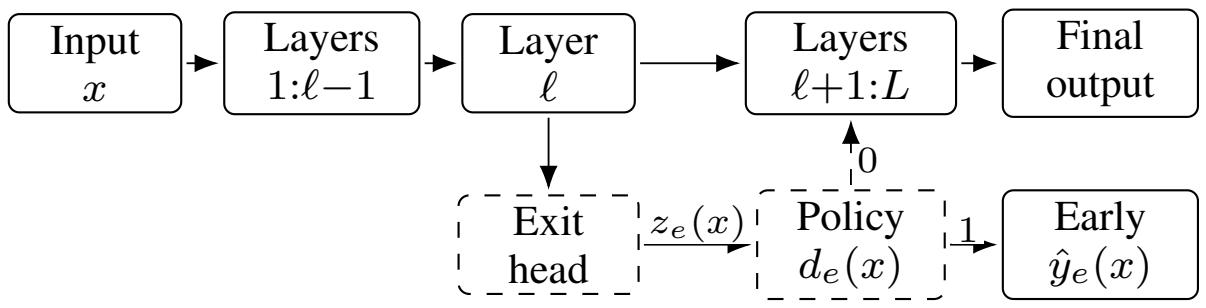
\begin{figure}[t]
\centering
\begin{tikzpicture}[
    >=Latex,
    font=\footnotesize,
    block/.style={
        draw,
        rounded corners=2pt,
        align=center,
        minimum height=6.5mm,
        text width=1.10cm,
        inner xsep=2pt,
        inner ysep=2pt,
        outer sep=0pt
    },
    exitblock/.style={
        block,
        dashed
    },
    edge/.style={
        ->,
        line width=0.35pt,
        shorten >=1pt,
        shorten <=1pt
    },
    control/.style={
        edge,
        dashed
    },
    lab/.style={
        font=\scriptsize,
        fill=white,
        inner xsep=1pt,
        inner ysep=0.25pt
    }
]

\node[block, text width=1.00cm] (input) at (0,0)
    {Input\\$x$};
\node[block, text width=1.15cm] (pre) at (1.50,0)
    {Layers\\$1{:}\ell{-}1$};
\node[block, text width=1.00cm] (layer) at (3.00,0)
    {Layer\\$\ell$};
\node[block, text width=1.25cm] (post) at (5.00,0)
    {Layers\\$\ell{+}1{:}L$};
\node[block, text width=1.15cm] (final) at (6.65,0)
    {Final\\output};

\node[exitblock, text width=1.05cm] (head) at (3.00,-1.20)
    {Exit\\head};
\node[exitblock, text width=1.10cm] (policy) at (5.00,-1.20)
    {Policy\\$d_e(x)$};
\node[block, text width=1.15cm] (earlyout) at (6.65,-1.20)
    {Early\\$\hat{y}_e(x)$};

\draw[edge] (input) -- (pre);
\draw[edge] (pre) -- (layer);
\draw[edge] (layer) -- (post);
\draw[edge] (post) -- (final);

\draw[edge] (layer.south) -- (head.north);

\draw[edge] (head.east)
    -- node[lab, above, pos=0.5] {$z_e(x)$}
    (policy.west);

\draw[edge] (policy.east)
    -- node[lab, above, pos=0.4] {$1$}
    (earlyout.west);

\draw[control] (policy.north)
    -- node[lab, right, pos=0.4] {$0$}
    (post.south);

\end{tikzpicture}
\caption{Representative intermediate exit $e$ after layer $\ell$. The exit
head produces class scores $z_e(x)$, which the policy maps to the routing
decision $d_e(x)$. If $d_e(x)=1$, inference returns the early prediction
$\hat{y}_e(x)$; otherwise it continues through layers $\ell{+}1{:}L$. This
structure may be repeated at selected layers, with inference terminating at
the earliest asserted exit.}
\label{fig:intermediate_exit}
\end{figure}

\subsection{Early Exit Neural Networks}
\label{subsec:ee_nets}

Early exit neural networks (EENNs) attach auxiliary classifiers to intermediate layers and terminate inference once a confidence condition is satisfied, reducing average inference cost. BranchyNet~\cite{branchy}, Shallow-Deep Networks (SDN)~\cite{sdn} and MSDNet~\cite{msd} attach intermediate classifiers and use confidence-based criteria such as entropy or maximum softmax probability, with joint training helping easy samples avoid later layers and reserve deeper computation for harder inputs. Simpler confidence policies can avoid full softmax computation. For example, maximum-logit thresholding is sensitive to common shifts in the logits, while a top-two margin requires top-two selection and a subtraction, so both retain explicit arithmetic decision logic.

On FPGA, EENNs create tension between data-dependent, per-sample control flow and statically scheduled dataflow pipelines. ATHEENA~\cite{ATHEENA} addresses this by extending fpgaConvNet with a control-and-dataflow graph (CDFG) that represents conditional execution. Its profiler estimates the probability $p$ that samples reach each downstream stage and uses this to size stage throughput and area for the expected early exit workload in each stage accordingly.

ATHEENA therefore requires softmax decision layers, conditional buffers, split/merge layers and floating-point computation. While these components enable dynamic execution, they add control logic and arithmetic overhead. Part of the resource and latency budget saved by early termination is consequently spent computing the confidence signal that controls it.

\section{Methodology}
\label{sec:methodology}

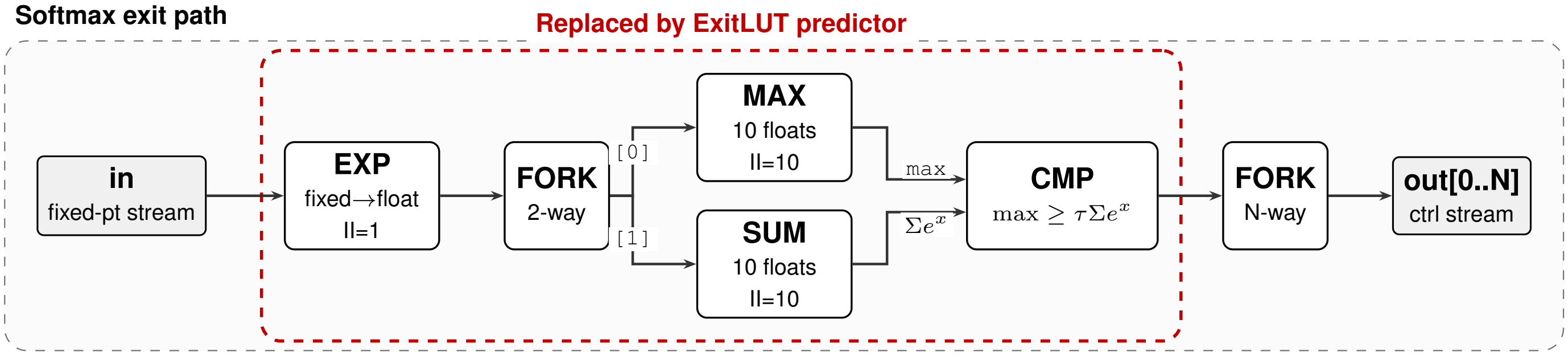
\begin{figure*}[t]
\centering
\resizebox{\textwidth}{!}{%
\begin{tikzpicture}[
  block/.style={
    draw=black, rectangle, rounded corners=3pt,
    minimum height=1.18cm, minimum width=1.7cm,
    align=center, font=\small\sffamily, fill=white, line width=0.6pt
  },
  port/.style={
    draw=black, rectangle, rounded corners=2pt,
    minimum height=0.85cm, minimum width=1.35cm,
    align=center, font=\small\sffamily, fill=gray!12, line width=0.55pt
  },
  wire/.style={-{Stealth[length=4pt,width=3pt]}, thick, draw=black!80},
  wlabel/.style={font=\scriptsize\ttfamily, fill=white, inner sep=1.2pt},
  sectbox/.style={
    draw=black!55, dashed, rounded corners=7pt,
    inner sep=10pt, fill=gray!3
  },
  absorbbox/.style={
    draw=red!75!black,
    dashed,
    rounded corners=6pt,
    line width=1.0pt,
    inner sep=7pt
  },
  boxlabel/.style={
    font=\footnotesize\bfseries\sffamily,
    text=red!75!black,
    fill=white,
    inner sep=1.5pt
  }
]

\node[port] (in) {\textbf{in}\\{\scriptsize fixed-pt stream}};

\node[block, right=0.85cm of in] (exp) {\textbf{EXP}\\{\scriptsize fixed$\to$float}\\{\scriptsize II=1}};

\node[block, right=0.7cm of exp, minimum width=1.15cm] (forki) {\textbf{FORK}\\{\scriptsize 2-way}};

\node[block, above right=-0.45cm and 0.95cm of forki] (max) {\textbf{MAX}\\{\scriptsize 10 floats}\\{\scriptsize II=10}};

\node[block, below right=-0.45cm and 0.95cm of forki] (sum) {\textbf{SUM}\\{\scriptsize 10 floats}\\{\scriptsize II=10}};

\coordinate (mid) at ($(max.east)!0.5!(sum.east)$);

\node[block, right=1.25cm of mid, minimum width=2.1cm] (cmp)
{\textbf{CMP}\\{\scriptsize $\max \geq \tau \Sigma e^x$}};

\node[block, right=0.7cm of cmp, minimum width=1.15cm] (forko)
{\textbf{FORK}\\{\scriptsize N-way}};

\node[port, right=0.7cm of forko] (out)
{\textbf{out[0..N]}\\{\scriptsize ctrl stream}};

\draw[wire] (in.east) -- (exp.west);

\draw[wire] (exp.east) -- (forki.west);

\draw[wire] (forki.east) -- ++(0.25,0) |- (max.west)
  node[wlabel, pos=0.22, above] {[0]};

\draw[wire] (forki.east) -- ++(0.25,0) |- (sum.west)
  node[wlabel, pos=0.22, below] {[1]};

\coordinate (cmp_hi) at ($(cmp.north west)!0.35!(cmp.south west)$);
\coordinate (cmp_lo) at ($(cmp.north west)!0.65!(cmp.south west)$);

\coordinate (rk1) at ($(max.east)+(0.38,0)$);
\coordinate (rk2) at (rk1 |- cmp_hi);

\coordinate (sk1) at ($(sum.east)+(0.38,0)$);
\coordinate (sk2) at (sk1 |- cmp_lo);

\draw[wire] (max.east) -- (rk1) -- (rk2) -- (cmp_hi);
\node[wlabel, above] at ($(rk2)!0.5!(cmp_hi)$) {max};

\draw[wire] (sum.east) -- (sk1) -- (sk2) -- (cmp_lo);
\node[wlabel, below] at ($(sk2)!0.5!(cmp_lo)$) {$\Sigma e^x$};

\draw[wire] (cmp.east) -- (forko.west);
\draw[wire] (forko.east) -- (out.west);

\begin{scope}[on background layer]
  \node[sectbox, fit=(in)(exp)(forki)(max)(sum)(cmp)(forko)(out)] (mainbox) {};
\end{scope}

\node[font=\footnotesize\bfseries\sffamily, anchor=south west]
  at (mainbox.north west) {Softmax exit path};

\begin{scope}[on background layer]
  \node[absorbbox, fit=(exp)(forki)(max)(sum)(cmp)] (absorbed) {};
\end{scope}

\node[boxlabel, anchor=south]
  at ($(absorbed.north)+(0,2pt)$)
  {Replaced by ExitLUT predictor};

\end{tikzpicture}%
}
\caption{ATHEENA exit-decision path. The red dashed region denotes the portion replaced by the ExitLUT predictor.}
\label{fig:lut_softmax_absorb}
\end{figure*}

Conceptually, an intermediate exit contains two components: an auxiliary classifier and an exit policy. The auxiliary classifier is attached after a selected hidden layer, as shown in Figure~\ref{fig:intermediate_exit}, and produces a vector of \(C\) class scores \(\mathbf{z}_e(x)\) for input \(x\). These scores determine the early class prediction. The exit policy uses the same scores to produce a routing decision \(d_e(x)\in\{0,1\}\), where \(d_e(x)=1\) terminates inference at exit \(e\), while \(d_e(x)=0\) forwards through the remaining layers.

Conventional early exit policies first reduce the class-score vector to a scalar confidence, and then compare it against a fixed threshold at runtime:
\begin{equation}
    d_e(x)
    =
    \ind
    \left[
        c_e\left(\mathbf{z}_e(x)\right)
        \geq \tau_e
    \right],
    \label{eq:generic_confidence_exit}
\end{equation}
where \(\ind[\cdot]\) is one when its condition is true and zero otherwise, \(c_e(\cdot)\) is a confidence function, such as maximum softmax probability or classification margin, and \(\tau_e\) is the corresponding floating-point exit threshold. Although the accelerator ultimately consumes only the one-bit routing decision \(d_e(x)\), evaluating this policy can require substantial arithmetic.

ExitLUT instead implements the required routing decision directly as a Boolean function of the finite-width scores visible to the FPGA. Let
\[
    \widetilde{\mathbf{z}}_e(x)
    =
    q_e\left(\mathbf{z}_e(x)\right)
    \in \mathcal{Q}_e
\]
denote the deployed representation of the exit scores, where \(q_e\) is the hardware quantization function and \(\mathcal{Q}_e\) is the finite set of possible quantized score vectors. ExitLUT computes
\begin{equation}
    d_e(x)
    =
    b_e\left(\widetilde{\mathbf{z}}_e(x)\right),
    \qquad
    b_e:\mathcal{Q}_e\rightarrow\{0,1\}.
    \label{eq:exitlut_boolean_exit}
\end{equation}
The Boolean function therefore produces the control signal required by the accelerator without explicitly evaluating the intermediate scalar confidence \(c_e(\mathbf{z}_e(x))\).

When routing and class selection can be fused, the Boolean function instead emits one exit flag per class:
\begin{equation}
    \mathbf{m}_e(x)
    =
    B_e\left(\widetilde{\mathbf{z}}_e(x)\right),
    \qquad
    B_e:\mathcal{Q}_e\rightarrow\{0,1\}^{C}.
    \label{eq:exitlut_class_flags}
\end{equation}
Here, \(m_{e,c}(x)=1\) indicates that inference should terminate at exit \(e\) with class \(c\). Any asserted flag triggers an exit, while the handling of multiple asserted flags is defined in Section~III-B. The abstraction is therefore the direct implementation of the required Boolean control relation, rather than the approximation of softmax or another arithmetic confidence score.

We realize this abstraction in two ways. For ATHEENA, the Boolean function is placed after the exit head as a compact LUT-based predictor trained to reproduce its thresholded softmax policy from quantized logits. For LUT-based neural networks, adding such a predictor would introduce unnecessary logic depth, so the class-specific Boolean conditions are folded directly into the exit L-LUT truth tables. These represent two placements of the same hardware abstraction: external Boolean policy replacement and embedded Boolean exits.

\subsection{Softmax-Policy Replacement}
\label{subsec:lut_exit_decision}

For ATHEENA, the exit scores \(\mathbf{z}_e(x)\in\mathbb{R}^{C}\) are the early exit logits, where \(C=10\) for MNIST and CIFAR-10. The baseline branch controller computes confidence as the maximum softmax probability,
\[
    c_e(\mathbf{z}_e(x))
    =
    \max_{c \in \{0,\ldots,C-1\}}
    \mathrm{softmax}(\mathbf{z}_e(x))_c .
\]
It then exits when this confidence exceeds the threshold:
\begin{equation}
    d_e(x)
    =
    \ind
    \left[
        c_e(\mathbf{z}_e(x))
        \geq \tau_e
    \right].
    \label{eq:softmax_exit_rule}
\end{equation}
In hardware, this requires exponentiation, maximum, threshold scaling and comparison. The ATHEENA softmax comparator is therefore realized as a multi-stage HLS pipeline, shown in Figure~\ref{fig:lut_softmax_absorb}, with a latency of 180 cycles~\cite{ATHEENA}.

ExitLUT replaces this arithmetic confidence path with a learned Boolean predictor:
\begin{equation}
    b_{\tau}
    \left(
        q_e(\mathbf{z}_e(x))
    \right)
    \approx
    \ind
    \left[
        c_e(\mathbf{z}_e(x))
        \geq \tau_e
    \right],
    \label{eq:atheena_boolean_exit}
\end{equation}
where $c_e(\mathbf{z}_e(x))$ is the maximum-softmax confidence.

This modifies the accelerator dataflow presented to the toolflow by replacing the arithmetic softmax-confidence block with the LUT-based Boolean predictor. After this replacement, the standard DSE as described in Section~\ref{subsec:fpga_nn} is rerun, allowing resources no longer required by the softmax decision path to be reallocated across the remaining dataflow datapath.

\subsubsection{Training data generation}
To match the behavior of the original ATHEENA early exit policy, the LUT network is trained using logits collected from the ATHEENA model. The branchy network is first evaluated on the training split, and the early exit logit vectors are saved together with their corresponding softmax exit labels to form training and validation splits. For each training example $i$, the saved dataset contains
\begin{equation}
    \left(
        \mathbf{z}^{(i)},
        y^{(i)}
    \right),
    \qquad
    y^{(i)}
    =
    \ind
    \left[
        c_e
        \left(
            \mathbf{z}^{(i)}
        \right)
        \geq \tau
    \right].
\end{equation}
Labels are generated from the floating-point ONNX logits, preserving the original softmax-based exit policy. The LUT predictor is then trained on the quantized logit representation used in hardware. Consequently, samples near the threshold may collapse to identical quantized inputs; any residual policy mismatch is captured by the predictor validation accuracy.

\subsubsection{Input representation}
The ATHEENA datapath represents early exit logits using \texttt{ap\_fixed<16,8>}, i.e. 16 total bits with 8 integer bits including the sign bit. We pack the sequential AXI-streamed logits into the LUT-network input word by extracting the signed integer byte from each 16-bit value. For logit component $z_c$, this is equivalent to
\begin{equation}
    q_c = \mathrm{clip}\!\left(\left\lfloor z_c \right\rfloor, -128, 127\right),
    \quad q_c \in \{-128,\ldots,127\}.
    \label{eq:logit_quantisation}
\end{equation}
Thus, the eight fractional bits are discarded and the remaining signed byte is used as the LUT-network input.

The main source of disagreement with the floating-point baseline arises from this deliberate truncation of \texttt{ap\_fixed<16,8>} logits to 8-bit integer addresses before LUT evaluation. Samples close to the threshold boundary can therefore change the decision after quantization. We therefore train and validate the LUT predictor using these truncated inputs, while retaining the original floating-point softmax decision as the supervision target.

\subsubsection{Training objective}
The LUT exit predictor is trained as a binary classifier on the collected logit dataset. For each sample, the floating-point early exit logits $\mathbf{z}^{(i)}$ are labelled using the original softmax confidence rule, while the predictor input is produced using the quantization path $q_e$ used in hardware. Given labels $y^{(i)}$ and quantized inputs $q_e(\mathbf{z}^{(i)})$, the predictor $g_{\theta}$ is optimized using cross-entropy loss:
\begin{equation}
    \theta^{\star}
    =
    \arg\min_{\theta}
    \sum_i
    \mathcal{L}_{\mathrm{CE}}
    \left(
        g_{\theta}
        \left(
            q_e
            \left(
                \mathbf{z}^{(i)}
            \right)
        \right),
        y^{(i)}
    \right).
    \label{eq:lut_training_loss}
\end{equation}

We first train the network fully connected to derive sparse connectivity masks, before training the sparse model~\cite{assemble}. We select the checkpoint with highest validation exit-decision accuracy and convert each quantized neuron to a deployable LUT by exhaustive truth-table enumeration~\cite{logicnets}.

\subsubsection{ATHEENA integration}
The generated Verilog is integrated into ATHEENA as shown in Figure~\ref{fig:lut_softmax_absorb}. At runtime, the ten logits arrive sequentially, one byte per cycle, and are accumulated over ten cycles. Once all ten bytes arrive, they are passed in parallel to the LUT network. The first stage contains one input L-LUT per logit byte, so each logit is quantized independently in parallel. The resulting low-bit activations are then processed in two cycles by the remaining sparse L-LUT layers to produce the single-bit exit decision. This output replaces the original softmax-based branch decision, while the rest of the ATHEENA optimizer and datapath mapping remain unchanged. The resulting exit-decision latency is reduced from 180 to 12 cycles, with the floating-point softmax path replaced entirely by DSP-free LUT logic.

\subsubsection{LUT-network architecture}
The LUT predictor follows the NeuraLUT-Assemble structure. Its input stage contains one L-LUT per logit byte. Each input L-LUT folds batch normalization, learned scalar bias $b$ and a learned-scale activation quantizer for that feature into a single lookup~\cite{assemble,cascade}:
\begin{equation}
    Q_c(q_c) = \mathrm{Quant}\left(\mathrm{BN}_c(q_c) + b\right)
    \label{eq:layerq}
\end{equation}
These ten lookups are evaluated in parallel following the accumulation. Since input quantization is implemented in LUT fabric, it introduces no DSPs or floating-point operators.

After input quantization, the exit network is implemented as a sequence of LUT-based quantized layers with fixed sparse connectivity. We implement six layers, the first of which has sixty-four neurons. Each subsequent layer has half the number of neurons as the previous layer. Each neuron has a fixed sparsity fan-in of two and uses 3-bit activations, except for the first layer, which uses 4-bit activations to preserve additional input precision. The network produces two 3-bit output scores, corresponding to the continue and exit classes. A final LUT implements the comparison
between the two scores.
The deployed network therefore emits the branch-control bit required by ATHEENA directly, without requiring an auxiliary softmax or arithmetic confidence-comparison stage. This architecture is fixed across all evaluated ATHEENA models.

\subsection{Embedded Exits}

For LUT-based neural networks, instantiating $b_e$ as a separate LUT predictor after each intermediate layer would add LUT layers and pipeline stages. This is undesirable because the base network is already implemented as shallow combinational logic between registers. We therefore use the fused form in Eq.~\ref{eq:exitlut_class_flags}, where the exit decision and class prediction are represented by class-specific Boolean flags.

Exits are inserted after hidden layers immediately before pipeline registers, so an asserted exit can reach a decision before later pipeline stages are executed. Let $s_{e,c}(x)$ denote the quantized output score for class $c$ at exit $e$. Each exit is assigned an integer threshold vector $\mathbf{T}_e \in \mathbb{Z}^{C}$ and emits
\begin{equation}
    m_{e,c}(x)
    =
    \ind
    \left[
        s_{e,c}(x)
        >
        \mathbf{T}_{e,c}
    \right],
    \qquad
    c \in \{0,\ldots,C-1\}.
    \label{eq:class_threshold_flags}
\end{equation}
Here $\mathbf{T}_{e,c}$ is an integer threshold applied to quantized class scores, in contrast to the floating-point confidence threshold $\tau$ used by conventional softmax-based exits. Thresholds are initialized from the maximum non-true-class score, so using a strict inequality prevents those boundary cases from triggering an early exit.

The branch decision is then
\begin{equation}
    d_e(x)
    =
    \bigvee_{c=0}^{C-1}
    m_{e,c}(x).
    \label{eq:lut_exit_or}
\end{equation}
If any bit is asserted, inference completes at exit $e$. If multiple class flags are asserted at the same exit, a fixed priority rule selects the lowest-index asserted class; if no bit is asserted, execution continues to the next pipeline stage. If no embedded exit fires, the final classifier is evaluated normally. These threshold tests are folded into the generated L-LUTs, so no separate confidence module, softmax block or arithmetic thresholding stage is required. For LUT-based networks, ExitLUT therefore reduces average time-to-decision by allowing valid class decisions to be emitted at intermediate pipeline stages. Since the datapath remains spatially instantiated, resource usage is not expected to decrease; the benefit is lower average latency with near-zero area overhead.

\subsubsection{Early Exit Training Methodology}
\label{subsec:ee-training}

We attach auxiliary output layers after selected hidden layers, immediately before the pipeline registers, and train following NeuraLUT-Assemble~\cite{assemble}. A hidden layer produces the intermediate activation, an exit layer produces quantized class scores and the following register marks the hardware stage boundary where inference may terminate. Sparse exit connections are learned after dense training. Training uses PyTorch with Stochastic Gradient Descent with Warm Restarts~\cite{sgd} and Decoupled Weight Decay~\cite{decoupled}. Each hidden sub-network uses batch normalization and Brevitas-based quantization~\cite{brevitas}.

The training objective is augmented with weighted cross-entropy losses at each early exit~\cite{deep-nets}. For exits $k < K$, the loss is scaled by $\lambda_k$, while the final classifier $K$ remains unscaled:
\begin{equation}
\label{eq:obj}
\mathcal{L}_{\text{total}}
=
\sum_{k=1}^{K-1} \lambda_k \, \mathrm{CE}\!\big(f_k(x;\theta), y\big)
\;+\;
\mathrm{CE}\!\big(f_K(x;\theta), y\big).
\end{equation}
The weight $\lambda_k$ controls the strength of auxiliary supervision at the exits, biasing capacity toward intermediate representations and trading a little accuracy for reduced inference depth.

\subsubsection{Hardware-Aware Exits}

\begin{figure}[b]
  \centering
  \includegraphics[width=\linewidth]{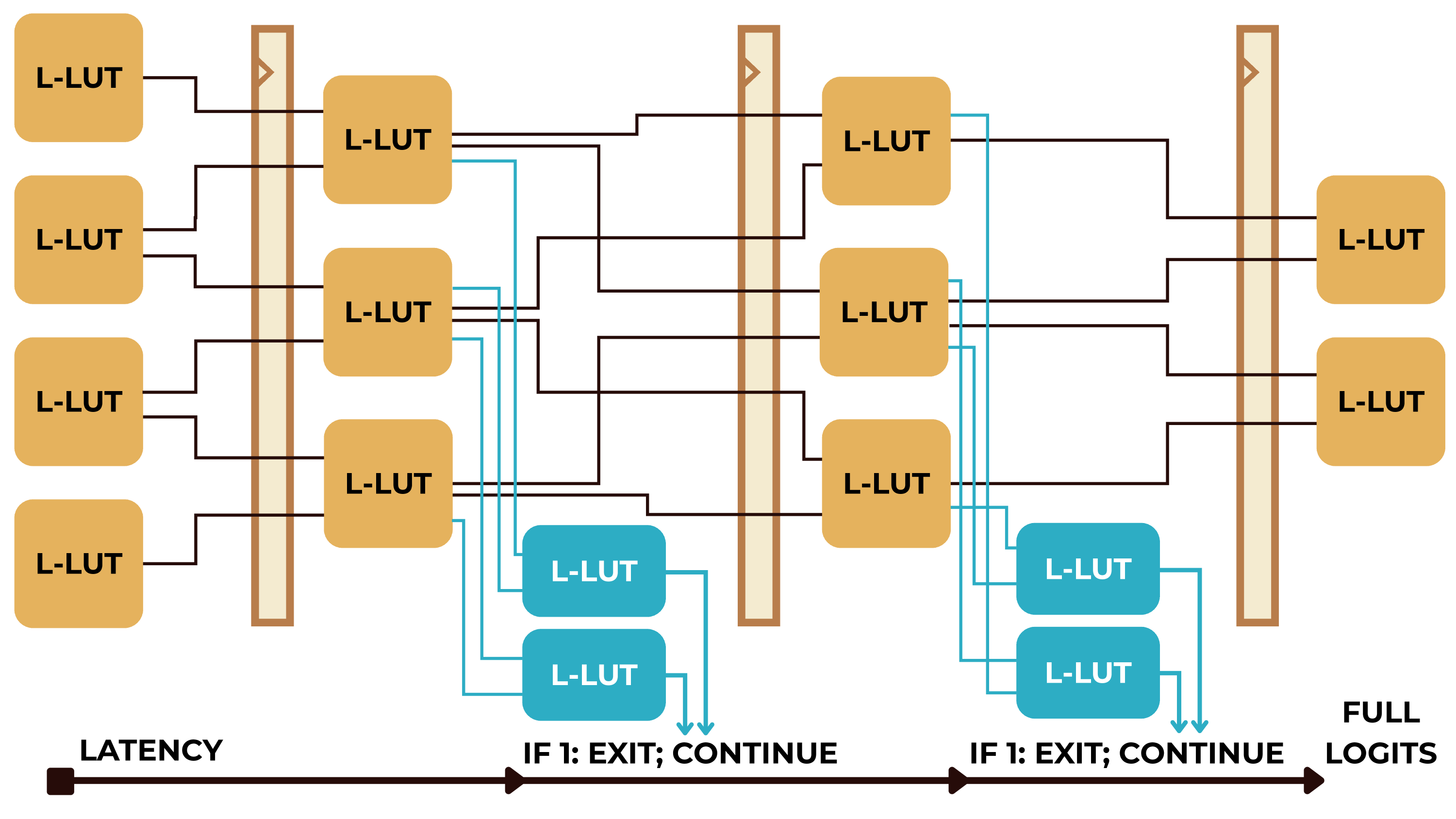}
  \caption{Embedded ExitLUT exits in a pipelined LUT-based network. Class-specific threshold tests are folded into the exit L-LUTs, so asserted flags terminate inference without a separate confidence module.}
  \label{fig:lut-diagram}
\end{figure}

To enable early exits without disrupting the low-latency datapath of LUT-based NNs, each exit compares its quantized output scores against fixed per-class integer thresholds. Conceptually, each class performs a fixed integer threshold test, but during LUT enumeration this comparison is folded into the corresponding L-LUT so that the exit emits Boolean class flags directly.

For an input $x$, let $D_{\mathbf{T}}(x)$ denote the index of the exit at which inference terminates under threshold set $\mathbf{T}=\{\mathbf{T}_e\}_e$, and let $\mathcal{D}$ be the training dataset. Let $\mathrm{Acc}(\mathbf{T})$ be the classification accuracy induced by $\mathbf{T}$ and let $\mathrm{Acc}_0$ be the accuracy of the full model without early exits. We seek integer thresholds that reduce expected exit depth, a hardware-faithful proxy for latency, subject to a bounded accuracy loss:
\begin{equation}
\min_{\mathbf{T}}\; \mathbb{E}_{x\sim\mathcal{D}}[D_{\mathbf{T}}(x)]
\quad\text{s.t.}\quad
\mathrm{Acc}(\mathbf{T}) \ge \mathrm{Acc}_0 - \Delta_{\max}.
\label{eq:acc-constr}
\end{equation}

Thresholds are initialized using training-set inference. For each exit and class, we record the maximum score observed when that class is not the true label. Since exits use the strict test in Eq.~\ref{eq:class_threshold_flags}, these initial thresholds \(\mathbf{T}^{(0)}_{e,c}\) admit no false early exits on the training set. Starting from these conservative bounds, thresholds are monotonically lowered to encourage earlier exiting. No threshold selection uses the test set.

At each exit $e$, we parameterize threshold adjustment by a single non-negative integer reduction $\Gamma_e$ applied uniformly across all classes at that exit. Thus, each candidate threshold is \(\mathbf{T}_{e,c}=\mathbf{T}^{(0)}_{e,c}-\Gamma_e\) for class \(c\). This preserves relative inter-class margins and reduces the search at each exit to a single discrete parameter. This sweep is a heuristic approximation to Eq.~\ref{eq:acc-constr}, rather than a global solution. For a given $\Gamma_e$, candidate thresholds are evaluated on the training set to estimate both expected accuracy loss and average exit depth. We sweep $\Gamma_e$ from zero, terminating when Eq.~\ref{eq:acc-constr} is first violated, and select the last feasible value before the first observed violation.

Exits are optimized in order from the earliest to the latest pipeline stage, so that threshold reductions are first allocated to exits that can terminate inference with the smallest executed pipeline depth. This ordering prioritizes latency reduction while enforcing the specified accuracy constraint.

Each exit-output L-LUT originally emits a quantized class score. ExitLUT replaces this output truth table with a binary truth table whose entry is one if the original quantized output is strictly greater than the optimized class threshold. Therefore, threshold comparison is performed at truth-table generation time rather than by a runtime comparator. Any asserted bit generates an exit-valid signal and encodes a candidate predicted class, eliminating the separate confidence-evaluation pipeline.

\section{Softmax-Policy Replacement Evaluation}
\label{sec:atheena-eval}

\subsection{Experimental Setup}
\label{subsec:atheena-setup}

\begin{table}[t] 
\centering 
\caption{ZC706 Branchy-LeNet resource and throughput comparison showing the effect of replacing the softmax exit path and the effect of rerunning DSE. * indicates the limiting resource.}
\label{tab:resource_comparison} 
\scriptsize 
\tighttable 
\begin{tabular}{@{}lrrrrrr@{}} 
\toprule 
Design & LUT & FF & DSP & BRAM-18 & \makecell{Lim.\ Res.\\(\%)} & \makecell{Thru.\\(Samples/s, Gain)} \\ 
\midrule 

fpgaConvNet-1
& 75513* & 61361 & 295 & 55 & 35 & 13513, 1.00$\times$ \\ 

ATHEENA-1
& 68383* & 63170 & 169 & 206 & 31 & 19434, 1.44$\times$ \\ 

\textbf{ExitLUT-1}
& 66419* & 60223 & 157 & 194 & 30 & 19442, 1.44$\times$ \\ 

\textbf{ExitLUT-opt-1}
& 81076* & 71433 & 275 & 198 & 37 & \textbf{34722, 2.57$\times$} \\ 

\midrule 

fpgaConvNet-2
& 105451 & 84761 & 470* & 89 & 52 & 21276, 1.00$\times$ \\ 

ATHEENA-2
& 128940* & 117138 & 407 & 239 & 59 & 47583, 2.24$\times$ \\ 

\textbf{ExitLUT-2}
& 126976* & 114191 & 395 & 227 & 58 & 47584, 2.24$\times$ \\ 

\textbf{ExitLUT-opt-2}
& 132717* & 117337 & 480 & 266 & 60 & \textbf{86805, 4.08$\times$} \\ 

\midrule 

fpgaConvNet-3
& 138194 & 120764 & 885* & 170 & 98 & 43384, 1.00$\times$ \\ 

ATHEENA-3
& 163637 & 142913 & 742* & 265 & 82 & 94170, 2.17$\times$ \\ 

\textbf{ExitLUT-3}
& 161673 & 139966 & 730* & 253 & 81 & 94173, 2.17$\times$ \\ 

\textbf{ExitLUT-opt-3}
& 169827* & 156875 & 644 & 317 & 78 & \textbf{96451, 2.22$\times$} \\ 

\bottomrule 
\end{tabular} 
\end{table}

\begin{table}[t]
\centering
\caption{Early exit decision-module resources. Percentages are relative to the corresponding full design in Table~\ref{tab:resource_comparison}.}
\label{tab:ee_overhead}
\scriptsize
\tighttable
\setlength{\tabcolsep}{2.1pt}
\begin{tabular}{@{}lrrrrrrrr@{}}
\toprule
Design & LUT & \% & FF & \% & DSP & \% & BRAM-18 & \% \\
\midrule

ATHEENA-1
& 3205 & 4.7 & 3407 & 5.4 & 12 & 7.1 & 12 & 5.8 \\

ExitLUT-1
& 1241 & 1.9 & 460 & 0.8 & 0 & 0.0 & 0 & 0.0 \\

ExitLUT-opt-1
& 1241 & 1.5 & 460 & 0.6 & 0 & 0.0 & 0 & 0.0 \\

\midrule

ATHEENA-2
& 3205 & 2.5 & 3407 & 2.9 & 12 & 2.9 & 12 & 5.0 \\

ExitLUT-2
& 1241 & 1.0 & 460 & 0.4 & 0 & 0.0 & 0 & 0.0 \\

ExitLUT-opt-2
& 1241 & 0.9 & 460 & 0.4 & 0 & 0.0 & 0 & 0.0 \\

\midrule

ATHEENA-3
& 3205 & 2.0 & 3407 & 2.4 & 12 & 1.6 & 12 & 4.5 \\

ExitLUT-3
& 1241 & 0.8 & 460 & 0.3 & 0 & 0.0 & 0 & 0.0 \\

ExitLUT-opt-3
& 1241 & 0.7 & 460 & 0.3 & 0 & 0.0 & 0 & 0.0 \\

\bottomrule
\end{tabular}
\end{table}

\begin{table}[t]
\centering
\caption{Accuracy comparison with Quantized LeNet. Per-Sample Agreement is with the softmax decision in ATHEENA.}
\label{tab:branchynet_comparison}
\scriptsize
\tighttable
\begin{tabular}{lrrrr}
\toprule
Network & \makecell{Top-1\\Acc. (\%)} & $p$ (\%) & $q$ (\%) & \makecell{Per-Sample Routing\\Agreement (\%)} \\
\midrule
fpgaConvNet LeNet  & 98.84 & -- & -- & --  \\
ATHEENA B-LeNet & 98.88 & 25.0 & 15.9 & \textit{100.0} \\
ATHEENA-ExitLUT-opt B-LeNet & 98.87 & 25.0 & 15.5 & 96.5 \\
\bottomrule
\end{tabular}
\end{table}

\begin{table}[t]
\centering
\caption{Throughput comparison of ATHEENA-ExitLUT, fpgaConvNet and ATHEENA two-stage accelerators on Xilinx VU440.}
\label{tab:athena_results}
\scriptsize
\setlength{\tabcolsep}{3pt}
\begin{tabular}{l l l c c r}
\hline
\textbf{Network (Task)} & \textbf{Toolflow} & \textbf{Lim.} & \textbf{Lim.} & \textbf{$p$ (\%)} & \textbf{Throughput} \\
 &  & \textbf{Res.} & \textbf{Res. (\%)} & & \textbf{(Samples/s, Gain)} \\
\hline
\textit{Triple Wins}~\cite{triplewins} & fpgaConvNet & DSP & 86 & --- & 19524, 1.00$\times$ \\
(MNIST~\cite{mnist})              & ATHEENA   & DSP & 81 & 25 & 54220, 2.78$\times$ \\
 & \textbf{ExitLUT-opt} & DSP & 88 & 25 & \textbf{72128, 3.69$\times$} \\

\hline
\textit{B-AlexNet}~\cite{branchy} & fpgaConvNet & DSP & 84 & --- & 8676, 1.00$\times$ \\
(CIFAR-10~\cite{cifar10})          & ATHEENA   & DSP & 88 & 34 & 17357, 2.00$\times$ \\
 & \textbf{ExitLUT-opt} & DSP & 89 & 34 & \textbf{19532, 2.25$\times$} \\
\hline
\end{tabular}
\end{table}

We follow the ATHEENA evaluation methodology to compare against fpgaConvNet and ATHEENA~\cite{ATHEENA}. Networks are trained in PyTorch, exported to ONNX, parsed by fpgaConvNet into a streaming dataflow architecture, and extended by ATHEENA with conditional execution. Hardware mappings are generated using the fpgaConvNet/ATHEENA DSE, followed by Vivado HLS 2019.1 synthesis and Vivado IP Integrator stitching. All designs use the same DMA-based host--FPGA interface, are conservatively clocked at 125\,MHz and report throughput over batches of 1024 samples.

The ZC706 Branchy-LeNet experiment separates the local cost of exit-policy replacement from the effect of re-optimizing the complete accelerator. Tables~\ref{tab:resource_comparison}, ~\ref{tab:ee_overhead} and ~\ref{tab:athena_results} use label where \textit{fpgaConvNet-$i$} denotes the static baseline, \textit{ATHEENA-$i$} the corresponding softmax-based early exit design, \textit{ExitLUT-$i$} the replacement-only design at the same datapath mapping and \textit{ExitLUT-opt-$i$} the result after rerunning DSE. Each DSE configuration is run ten times with independent simulated-annealing seeds.
For early exit designs, $p$ denotes the profiled continuation probability supplied to DSE, while $q$ denotes the measured fraction of samples continuing during evaluation.

We report post-quantization MNIST test accuracy for both LeNet and B-LeNet on the official benchmark~\cite{mnist}; continuation probability and sample-wise agreement with the original ATHEENA softmax decision are reported only for B-LeNet.

\def \baselineGraphColour {red}
\definecolor{combGraphColour}{RGB}{154,87,255}
\definecolor{exitlutGraphColour}{RGB}{0,128,0}

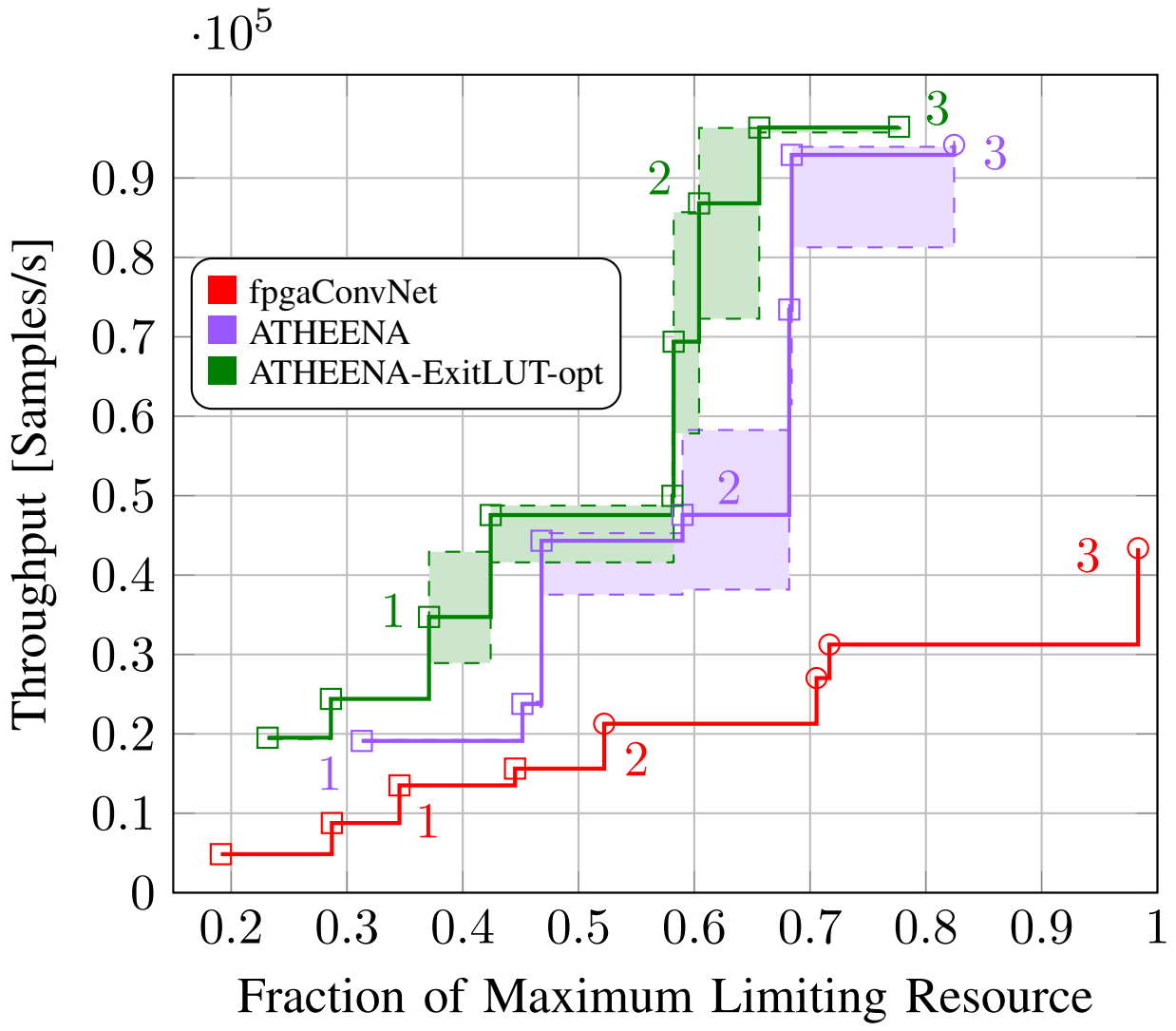
\begin{figure}[t]
    \centering
    \begin{tikzpicture}
        \begin{axis}[
            xlabel={Fraction of Maximum Limiting Resource},
            ylabel={Throughput [Samples/s]},
            xmin=0.15, xmax=1.00,
            ymin=0, ymax=103000,
            xtick={0.2,0.3,0.4,0.5,0.6,0.7,0.8,0.9,1.0},
            ytick={0,10000,20000,30000,40000,50000,60000,70000,80000, 90000, 1000000},
            legend pos=north west,
            ymajorgrids=true,
            xmajorgrids=true,
        ]
            \addplot[const plot,color=\baselineGraphColour,thick] 
                table[col sep=comma]
                {data_hw/lenet_se_baseline_HW.csv};

            \addplot[color=\baselineGraphColour,mark=o,only marks,]
                table[col sep=comma]{data_hw/lenet_se_baseline_HW_DSPS.csv};

            \addplot[color=\baselineGraphColour,mark=square,only marks,]
                table[col sep=comma]{data_hw/lenet_se_baseline_HW_LUTS.csv};

            \addplot[const plot,color=combGraphColour,thick]
                table[col sep=comma]
                {data_comb/comb_HW.csv};

            \addplot[color=combGraphColour,mark=o,only marks,]
                table[col sep=comma]{data_comb/comb_DSP_HW.csv};

            \addplot[color=combGraphColour,mark=square,only marks,]
                table[col sep=comma]{data_comb/comb_LUT_HW.csv};

            \addplot+[const plot,name path=20, color=combGraphColour,no markers,dashed] %
                table[col sep=comma]
                {data_comb/comb_upper_HW.csv};

            \addplot+[const plot,name path=30, color=combGraphColour,no markers,dashed] %
                table[col sep=comma]
                {data_comb/comb_lower_HW.csv};

            \addplot[combGraphColour!20] fill between[of=20 and 30];

            \addplot[const plot,color=exitlutGraphColour,thick]
                table[col sep=comma]
                {results-exitlut-hw/comb_HW.csv};

            \addplot[color=exitlutGraphColour,mark=square,only marks,]
                table[col sep=comma]{results-exitlut-hw/comb_LUT_HW.csv};

            \addplot+[const plot,name path=20, color=exitlutGraphColour,no markers,dashed] %
                table[col sep=comma]
                {results-exitlut-hw/comb_upper_HW.csv};

            \addplot+[const plot,name path=30, color=exitlutGraphColour,no markers,dashed] %
                table[col sep=comma]
                {results-exitlut-hw/comb_lower_HW.csv};

            \addplot[exitlutGraphColour!20] fill between[of=20 and 30];

            \node[color=combGraphColour] (pnt_a) at (axis cs:0.285,15000){$1$};

            \node[color=combGraphColour] (pnt_b) at (axis cs:0.63,51000){$2$};

            \node[color=combGraphColour] (pnt_c) at (axis cs:0.86,93200){$3$};

            \node[color=\baselineGraphColour] (pnt_d) at (axis cs:0.37,9000){$1$};

            \node[color=\baselineGraphColour] (pnt_e) at (axis cs:0.55,16800){$2$};

            \node[color=\baselineGraphColour] (pnt_f) at (axis cs:0.94,42500){$3$};

            \node[color=exitlutGraphColour] (pnt_g) at (axis cs:0.34,35500){$1$};

            \node[color=exitlutGraphColour] (pnt_h) at (axis cs:0.57,90000){$2$};

            \node[color=exitlutGraphColour] (pnt_i) at (axis cs:0.81,98800){$3$};

            \node[
                draw,
                fill=white,
                rounded corners,
                font=\scriptsize,
                anchor=north west,
                align=left
            ] at (axis cs:0.165,80000) {
                \textcolor{\baselineGraphColour}{\rule{0.8em}{0.8em}} fpgaConvNet\\
                \textcolor{combGraphColour}{\rule{0.8em}{0.8em}} ATHEENA\\
                \textcolor{exitlutGraphColour}{\rule{0.8em}{0.8em}} ATHEENA-ExitLUT-opt
            };
        
        \end{axis}
    \end{tikzpicture}

    \caption{Throughput--area results on the ZC706 at $p=25\%$, evaluated using test mixes with $q \in \{20\%,25\%,30\%\}$ (shaded, solid is $q=25\%$).}
    \label{area_thruput_optim_board_r}
\end{figure}

\subsection{Decision-Path Ablation}
\label{subsec:atheena-ablation}

Tables~\ref{tab:resource_comparison} and~\ref{tab:ee_overhead} show the effect of replacing the ATHEENA softmax decision path. The replacement-only \textit{ExitLUT-$i$} rows keep the ATHEENA datapath fixed, so their throughput remains essentially unchanged from \textit{ATHEENA-$i$}. The resource reduction is therefore attributable to the new exit controller itself rather than to a different accelerator mapping.

The LUT-based decision path uses 2.58$\times$ fewer exit-decision LUTs than ATHEENA, reduces the corresponding FF usage by 7.41$\times$ and removes all decision-path DSPs and BRAMs. Table~\ref{tab:branchynet_comparison} shows that ATHEENA-ExitLUT matches ATHEENA in top-1 accuracy and aggregate continuation probability, while agreeing with the softmax routing decision on 96.5\% of test samples. Of the remaining samples, 55.6\% are ExitLUT-only exits, for which ExitLUT exits while ATHEENA continues, and 44.4\% are ExitLUT-only continues, where ATHEENA exits while ExitLUT continues.

The ATHEENA softmax policy is a reference routing policy rather than an oracle for classification correctness. A disagreement therefore indicates that the two policies route a sample differently, not that ExitLUT necessarily makes an incorrect classification decision. Despite these differences, both policies achieve higher top-1 accuracy than the static baseline, and their continuation probabilities are nearly identical. They therefore induce essentially the same expected second-stage traffic and, for a fixed hardware mapping, the same average execution throughput. The throughput gain after re-optimization instead arises because the cheaper ExitLUT controller releases resources that DSE can reinvest in the datapath.

\subsection{Re-optimized Datapath Throughput}
\label{subsec:atheena-reoptimized}

The \textit{ExitLUT-opt-$i$} rows in Table~\ref{tab:resource_comparison} rerun DSE after the softmax path has been replaced. These designs use more total resources than the replacement-only \textit{ExitLUT-$i$} rows because the freed resource budget is reinvested into the accelerator datapath rather than left unused. Since the exit policy is unchanged, the accuracy, exit probability and sample-wise agreement reported in Table~\ref{tab:branchynet_comparison} characterize the early exit behavior. The difference is that DSE now chooses higher-throughput datapath folding factors and greater weight parallelism for the same design size.
The single-run throughput results from DSE are $23.2\pm2.6$ and $84.4\pm22.1$\,kFPS (mean$\pm$std) at the 25\% and 50\% resource budgets.

Figure~\ref{area_thruput_optim_board_r} plots the resulting throughput--area frontier at $p=25\%$, using test mixes with measured second-stage fractions $q\in\{20\%,25\%,30\%\}$. ATHEENA-ExitLUT shifts the implemented frontier above ATHEENA, with the largest improvements at lower resource fractions where the softmax exit path consumes a larger share of the limiting resource budget. 
The peak ZC706 point reaches $4.08\times$ fpgaConvNet and $1.82\times$ ATHEENA throughput because, in LUT-limited designs, freed softmax-path LUTs translate into greater datapath parallelism. 
DSE uses extra BRAM weight-bank replicas for concurrent MAC paths as required by increased convolution parallelism.
The smaller gain at the largest design point is consistent with the bottleneck moving toward the first stage before the exit, where reducing exit-controller cost cannot be converted into throughput as directly. However, here the advantage becomes a stronger second stage of the accelerator, which is able to maintain this high throughput as $q$ increases.

\subsection{Larger Two-Stage Accelerators}
\label{subsec:atheena-large}

Table~\ref{tab:athena_results} repeats the comparison on larger VU440 two-stage accelerators. Here, $p$ is the profiled continuation probability supplied to DSE. For both Triple-Wins~\cite{triplewins} and B-AlexNet~\cite{branchy}, ATHEENA and ATHEENA-ExitLUT-opt use the same profiled second-stage fraction, but ATHEENA-ExitLUT-opt achieves higher throughput. This confirms the same story at a larger scale: the gain is not produced by exiting more aggressively, but by removing arithmetic-heavy exit control and allowing the freed resources to support the main datapath.

\section{Embedded Exits Evaluation}
\label{sec:neuralut-eval}

\subsection{Experimental Setup}
\label{subsec:neuralut-setup}

{%
\setlength{\tabcolsep}{2pt}
\begin{table*}[t]
\centering
\scriptsize
\caption{NeuraLUT-Assemble-ExitLUT architectures and training parameters used across our evaluation. II is the initiation interval.}
\renewcommand{\arraystretch}{0.89}
\label{table:assemble-configs}
\begin{tabular}{lccccccccccccc}
\toprule
Dataset & $w_\ell$ & $w_e$ & $a_\ell$ & $a_e$ & $F$ & $F_e$ & $\beta$ & $\beta_e$ & $L$ & $N$ & $S$ & $\lambda_k$ & II\\
\midrule
MNIST &
$[2160,360,2160,360,60,10]$ &
$[10,10,10,10,10]$ &
$[0,1,0,1,1,1]$ & 0 &
$[6,6,6,6,6,6]$ &
6 &
$[1,1,1,1,1,6]$ &
6 &
2 & 64 & 2 & 0.85 & 1 \\
JSC CERNBox &
$[320,160,80,40,20,10,5]$ &
$[5,5,5,5,5,5]$ &
$[0,1,1,1,1,1,1]$ & 0 &
$[1,2,2,2,2,2,2]$ &
2 &
$[8,4,4,4,4,4,8]$ &
8 &
2 & 64 & 2 & 0.85 & 1 \\
JSC OpenML &
$[320,160,80,40,20,10,5]$ &
$[5,5,5,5,5,5]$ &
$[0,1,1,1,1,1,1]$ & 0 &
$[1,2,2,2,2,2,2]$ &
2 &
$[6,3,3,3,3,3,8]$ &
8 &
2 & 64 & 2 & 0.85 & 1 \\
\bottomrule
\end{tabular}
\end{table*}
}

\begin{figure*}[t]
\centering
\subfloat[JSC CERNBox]{%
  \centering
  \begin{tikzpicture}[scale=0.8]  
        \begin{axis}[
            xlabel={Test Accuracy},
            xlabel style={yshift=-3pt},
            ylabel={Area$\times$Delay},
            legend style={
                at={(0.02,1.9)},
                anchor=north west
            },
            grid=major,
            axis y line=left,
            axis x line=bottom,
            xmin=73.98, xmax=74.87,
            ymin=1.75e4, ymax=6.1e4,
            xscale=0.88,
            yscale=0.52
        ]
        
        \addplot[
            only marks,
            mark=*,
            color=blue
        ]
        coordinates {
        (74.02,1.93e4)
        (74.14,2.02e4)
        (74.24,2.08e4)
        (74.33,2.10e4)
        (74.44,2.22e4)
        (74.52,2.29e4)
        (74.62,2.50e4)
        (74.71,2.61e4)
        (74.82,3.75e4)
        };
        \addlegendentry{Early Exits}

        \addplot[
            only marks,
            mark=diamond*,
            mark size=3,
            color=red
        ]
        coordinates {
        (74.83,6.0e4)
        };
        \addlegendentry{Static Model}

        \end{axis}

        \end{tikzpicture}
}\hfill
\subfloat[JSC OpenML]{%
  \centering
  \begin{tikzpicture}[scale=0.8]  
        \begin{axis}[
            xlabel={Test Accuracy},
            xlabel style={yshift=-3pt},
            ylabel={Area$\times$Delay},
            legend style={
                at={(0.02,1.9)},
                anchor=north west
            },
            grid=major,
            axis y line=left,
            axis x line=bottom,
            xmin=75.03, xmax=75.9,
            ymin=3.3e3, ymax=1.3e4,
            xscale=0.88,
            yscale=0.52
        ]
        
        \addplot[
            only marks,
            mark=*,
            color=blue
        ]
        coordinates {
        (75.07,3.62e3)
        (75.18,3.95e3)
        (75.26,4.35e3)
        (75.34,4.68e3)
        (75.46,4.73e3)
        (75.55,4.88e3)
        (75.68,5.99e3)
        (75.77,6.14e3)
        (75.87,7.48e3)
        };
        \addlegendentry{Early Exits}

        \addplot[
            only marks,
            mark=diamond*,
            mark size=3,
            color=red
        ]
        coordinates {
        (75.87,1.21e4)
        };
        \addlegendentry{Static Model}

        \end{axis}

        \end{tikzpicture}
}\hfill
\subfloat[MNIST]{%
  \centering
  \begin{tikzpicture}[scale=0.8]  
        \begin{axis}[
            xlabel={Test Accuracy},
            xlabel style={yshift=-3pt},
            ylabel={Area$\times$Delay},
            legend style={
                at={(0.02,1.9)},
                anchor=north west
            },
            grid=major,
            axis y line=left,
            axis x line=bottom,
            xmin=96.95, xmax=97.85,
            ymin=5.8e3, ymax=3.4e4,
            xscale=0.88,
            yscale=0.52
        ]
        
        \addplot[
            only marks,
            mark=*,
            color=blue
        ]
        coordinates {
        (97.01,6.90e3)
        (97.12,7.64e3)
        (97.21,7.87e3)
        (97.30,8.26e3)
        (97.38,8.76e3)
        (97.51,1.16e4)
        (97.60,1.21e4)
        (97.71,1.44e4)
        (97.79,2.18e4)
        };
        \addlegendentry{Early Exits}

        \addplot[
            only marks,
            mark=diamond*,
            mark size=3,
            color=red
        ]
        coordinates {
        (97.80,3.3e4)
        };
        \addlegendentry{Static Model}

        \end{axis}

        \end{tikzpicture}
}
\caption{Test accuracy (\%) versus area$\times$delay ($\text{P-LUTs}\times\text{ns}$; lower is better) for P1 ExitLUT models across $\Delta_{\max} \in \{0.0\%,0.1\%,\ldots,1.0\%\}$.}
\label{fig:exiguity-graphs}
\end{figure*}
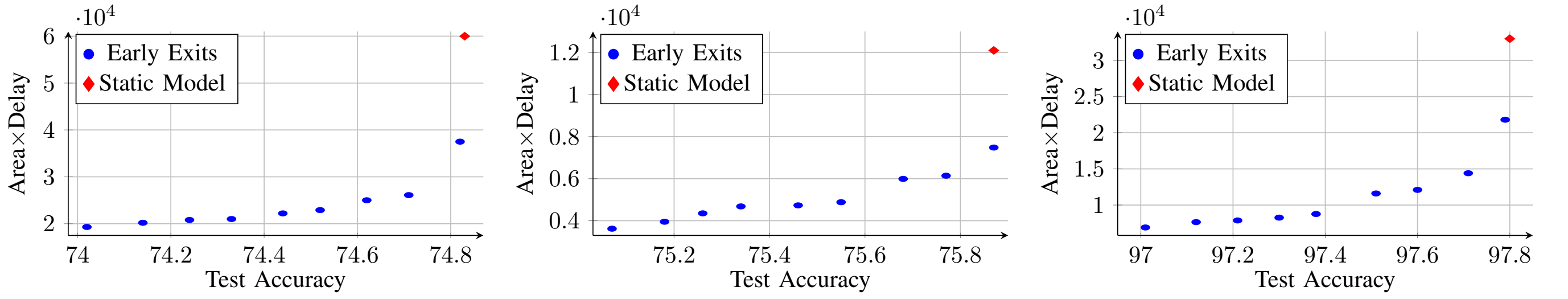

\begin{table}[t]
\centering
\caption{Out-of-context ExitLUT accuracy--latency trade-off. Average latency falls as accuracy constraints are relaxed.}
\label{tab:pipeline-latency}
\footnotesize
\setlength{\tabcolsep}{1.6pt}
\renewcommand{\arraystretch}{0.90}
\begin{tabular*}{\columnwidth}{@{}lc l@{\extracolsep{\fill}}rrrr@{}}
\toprule
\textbf{Dataset} &
\textbf{$F_{\max}$} &
\textbf{Metric} &
\textbf{Static} &
\multicolumn{3}{c}{\textbf{ExitLUT $\Delta_{\max}$}} \\
\cmidrule(lr){5-7}
&
\textbf{S/E} &
&
&
$\boldsymbol{0}$ &
$\boldsymbol{0.1}$ &
$\boldsymbol{0.3}$ \\
\midrule

\multicolumn{7}{@{}l}{\textit{P1: pipeline every L-LUT layer}} \\
MNIST
& 916/927
& Acc. (\%)          & 97.80 & 97.79 & 97.71 & 97.51 \\
&
& \textbf{Lat. (ns)} & \textbf{6.55} & \textbf{4.26} & \textbf{2.82} & \textbf{2.27} \\
\addlinespace[3pt]
JSC OpenML
& 1202/1214
& Acc. (\%)          & 75.87 & 75.87 & 75.77 & 75.55 \\
&
& \textbf{Lat. (ns)} & \textbf{5.82} & \textbf{4.04} & \textbf{3.32} & \textbf{2.64} \\
\addlinespace[3pt]
JSC CERNBox
& 992/988
& Acc. (\%)          & 74.83 & 74.82 & 74.71 & 74.52 \\
&
& \textbf{Lat. (ns)} & \textbf{7.06} & \textbf{4.36} & \textbf{3.01} & \textbf{2.66} \\

\midrule

\multicolumn{7}{@{}l}{\textit{P3: pipeline every three L-LUT layers}} \\
MNIST
& 866/887
& Acc. (\%)          & 97.80 & 97.80 & 97.70 & 97.48 \\
&
& \textbf{Lat. (ns)} & \textbf{2.31} & \textbf{2.04} & \textbf{1.46} & \textbf{1.36} \\
\addlinespace[3pt]
JSC OpenML
& 1151/1147
& Acc. (\%)          & 75.87 & 75.87 & 75.78 & 75.57 \\
&
& \textbf{Lat. (ns)} & \textbf{1.74} & \textbf{1.73} & \textbf{1.25} & \textbf{1.18} \\
\addlinespace[3pt]
JSC CERNBox
& 447/449
& Acc. (\%)          & 74.83 & 74.83 & 74.73 & 74.51 \\
&
& \textbf{Lat. (ns)} & \textbf{4.45} & \textbf{4.17} & \textbf{3.25} & \textbf{2.90} \\

\bottomrule
\end{tabular*}
\vspace{0.25em}

{\raggedright
\scriptsize
ExitLUT columns are indexed by the permitted accuracy loss $\Delta_{\max}$ in percent.
$F_{\max}$ is reported in MHz as static/early exit. Latency is the test-set average.
\par}
\vspace{-0.8em}
\end{table}

\begin{table}[t]
\centering
\scriptsize
\setlength{\tabcolsep}{2.5pt}
\renewcommand{\arraystretch}{1.0}
\caption{Out-of-Context evaluation of ExitLUT against previous works, with results taken from cited conference papers.}
\label{tab:neuralut-eval}
\begin{threeparttable}
\begin{tabular}{llrrrrr}
\toprule
Dataset & Model & Acc. (\%) & LUTs & Freq. (MHz) & Lat. (ns) & ADP \\
\midrule
MNIST & \textbf{NLA-ExitLUT} & 97.8 & 5107 & \textbf{887} & \textbf{1.5} & \textbf{7.64e3} \\
      & \textit{NLA-Softmax} & \textit{97.8} & \textit{6502} & \textit{880} & \textit{4.9} & \textit{3.19e4} \\
      & NLA~\cite{assemble} & \textbf{97.9} & 5070 & 863 & 2.1 & 1.06e4 \\
      & TreeLUT~\cite{treelut} & 96.6 & 4478 & 791 & 2.5 & 1.12e4 \\
      & DWN~\cite{dwn} & 97.8 & \textbf{2092} & 873 & 9.2 & 1.92e4 \\
\midrule
JSC OpenML & \textbf{NLA-ExitLUT} & 75.9 & \textbf{1780} & \textbf{1147} & \textbf{1.5} & \textbf{2.68e3} \\
      & NLA~\cite{assemble} & 76.0 & \textbf{1780} & 941 & 2.1 & 3.92e3 \\
      & TreeLUT~\cite{treelut} & 75.6 & 2334 & 735 & 2.7 & 6.03e3 \\
      & DWN~\cite{dwn} & \textbf{76.3} & 6302 & 695 & 14.4 & 9.07e4 \\
\midrule
JSC CERNBox & \textbf{NLA-ExitLUT} & 74.8 & 8597 & \textbf{988} & \textbf{3.0} & \textbf{2.61e4} \\
      & NLA~\cite{assemble} & \textbf{75.0} & \textbf{8539} & 352 & 5.7 & 4.87e4 \\
      & NeuraLUT~\cite{NeuraLUT} & \textbf{75.0} & 92357 & 368 & 14.0 & 1.29e6 \\
      & LogicNets~\cite{logicnets}            & 71.8 & 37931 & 427 & 13 & 4.93e5 \\
\bottomrule
\end{tabular}

\vspace{1pt}
\begin{tablenotes}[flushleft]
\scriptsize
\item[] All designs listed use 0 BRAM and 0 DSP resources. ExitLUT uses $\Delta_{\max}$ of 0.04$\%$ for MNIST (P3), 0.02$\%$ for OpenML (P3) and 0.08$\%$ for CERNBox (P1). Area-Delay Product (ADP) is measured in P-LUT$\times$ns.
\end{tablenotes}
\end{threeparttable}
\end{table}

We evaluate LUT-based NN exits by running inference directly on the synthesized Verilog circuit and measuring test-set accuracy. Synthesis uses Vivado 2020.1 targeting the \texttt{xcvu9p-flgb2104-2-i} FPGA out-of-context. We report P-LUT utilization, maximum frequency and time-to-decision latency. Latency is measured from input arrival to either an embedded early exit decision or the final logits, and includes only datapath logic and pipeline-register latency, as illustrated in Figure~\ref{fig:lut-diagram}. This is to match prior LUT-based NN evaluations as both measurements stop at raw output availability.

Table~\ref{table:assemble-configs} lists the three trained NeuraLUT-Assemble (NLA) backbones and their exit parameters. Layer sizes $w_\ell$ give the number of L-LUTs per layer, while $w_e$ gives the exit L-LUTs. Assemble masks $a_\ell$ indicate layers with fixed sparse inputs. The $a_e=0$ setting means no fixed Assemble mask is imposed on the exit layers; their sparse connections are learned after dense training. Fan-ins $F$ and bit-widths $\beta$ define the main datapath connectivity and quantization, while $(F_e,\beta_e)$ define the exit layers. \((L,N,S)\) are the NLA subnetwork parameters. We evaluate two pipeline granularities: P1 inserts a register after every L-LUT layer, while P3 inserts a register after every three L-LUT layers. The first model targets MNIST digit classification~\cite{mnist} with a 60k/10k train-test split. The other two target jet substructure classification using CERNBox~\cite{CERNBox_jsc} with an 800k/200k split and OpenML~\cite{openml_jsc} with a 664k/166k split.

\subsection{External Softmax-Policy Replacement}
\label{subsec:neuralut-external-policy}

Table~\ref{tab:neuralut-eval} includes an external softmax-policy replacement exit, \textit{NLA-Softmax}, on MNIST to show why the ATHEENA-style replacement is not sufficient for LUT-based NNs. Although this design preserves accuracy, the external policy network adds logic depth and pipeline stages, increasing latency from $2.1$\,ns for static NeuraLUT-Assemble to $4.9$\,ns. This depth was acceptable when replacing ATHEENA's 180-cycle softmax path, but it is too large when the baseline already operates at nanosecond-scale latency. The LUT-based setting therefore requires embedded exits, where the threshold tests are folded into the exit heads instead of implemented as a separate decision module beyond the exit heads.

\subsection{Embedded Exit Time-to-Decision}
\label{subsec:neuralut-time-to-decision}

Table~\ref{tab:pipeline-latency} shows that embedded ExitLUT exits reduce average time-to-decision across all datasets while preserving the LUT-based execution model. The maximum clock frequency remains similar between static and early exit designs, so the gain comes from reducing the average number of pipeline stages evaluated per input rather than changing the cycle time.

The largest reductions occur for P1 designs, where registers after every L-LUT layer provide fine-grained exit points. 
With no allowed accuracy loss, P1 reduces average latency, achieving 1.44--1.62$\times$ speedup over the corresponding static design.
At $\Delta_{\max}=0.3\%$, P1 achieves 2.20--2.89$\times$ speedup over the static design across all datasets.
Figure~\ref{fig:exiguity-graphs} shows the same trend as an accuracy--area-delay trade-off: relaxing the accuracy constraint allows more samples to exit earlier, reducing average latency and therefore area-delay product.

Coarser P3 pipelining gives fewer useful exit boundaries, so fewer samples can be classified prior to a substantial fraction of the pipeline. Even so, at $\Delta_{\max}=0.3\%$, P3 still achieves a $1.47$--$1.69\times$ speedup over the static design. The result confirms that embedded exits are most valuable when the hardware has enough pipeline boundaries for early decisions before waiting for later pipeline stages to produce a result.

While thresholds are selected on the training set, the test-set accuracy loss in Table~\ref{tab:pipeline-latency} remains close to the requested $\Delta_{\max}$ values across datasets. This suggests that embedded 1-bit class flags capture stable decision margins, rather than overfitting.

\subsection{Pipeline Granularity and Prior LUT-Based NNs}
\label{subsec:neuralut-prior-work}

Table~\ref{tab:neuralut-eval} compares selected low-loss ExitLUT Pareto points for each dataset against prior LUT-based NNs. ExitLUT remains close to NeuraLUT-Assemble in accuracy and area, with under $1\%$ LUT overhead and the expected small peak accuracy loss from Section~\ref{subsec:ee-training}, but improves latency and area-delay product. The headline gain is time-to-decision: across these configurations, ExitLUT gives a $1.41$--$1.89\times$ average latency speedup over static NeuraLUT-Assemble.

JSC CERNBox shows why this matters for high-throughput designs. The larger CERNBox L-LUT circuits benefit from finer pipelining because P1 raises $F_{\max}$ from $449$\,MHz in P3 to $988$\,MHz with embedded exits. Normally, this additional pipelining would increase the number of stages each sample must traverse. ExitLUT changes the trade-off: at $\Delta_{\max}=0.3\%$, the P1 CERNBox design reaches $2.66$\,ns average time-to-decision, below the static P3 latency of $4.45$\,ns while retaining the much higher clock frequency. Thus, for CERNBox, embedded exits allow the design to pipeline more aggressively for throughput while still lowering average decision latency.

\section{Conclusion and Future Work}
\label{sec:furtherwork}

This paper presented ExitLUT, a hardware-matched reformulation of early exit FPGA inference in which the continue-or-exit decision is implemented as Boolean LUT logic rather than arithmetic confidence evaluation. In ATHEENA-style dataflow CNN accelerators, replacing the softmax exit path reduces exit-decision LUT usage by 2.58$\times$, removes decision-path DSP usage and preserves the deployed exit behavior; after DSE reuses the freed resources, throughput improves by up to $1.82\times$ over ATHEENA. In LUT-based NNs, where a standalone predictor is too costly for the nanosecond-scale baseline, ExitLUT instead embeds class-specific threshold tests into L-LUT truth tables, achieving a $1.41$--$1.89\times$ average time-to-decision speedup with under $1\%$ LUT-area overhead. These results show that Boolean early exits are most effective when placed at the level matching the network backbone: external Boolean replacement for arithmetic dataflow accelerators, and embedded Boolean exits for LUT-based networks. Future work will evaluate larger-class models, where predictor complexity depends more on class count and routing-boundary complexity than backbone depth, alongside joint exit-placement and threshold optimization.

\bibliographystyle{IEEEtran}
\bibliography{bibliography}

@INPROCEEDINGS{cascade,
  author={Cassidy, Oliver and Andronic, Marta and Constantinides, George A.},
  booktitle={2026 36th International Conference on Field-Programmable Logic and Applications (FPL)}, 
  title={{CascadeLUT: Information-Ordered Streaming Inference for Bandwidth-Constrained FPGAs}}, 
  year={2026},
  volume={},
  number={},
  pages={116-121},
  doi={10.23919/FPLCPS00026.2026.00028}}

@article{nns,
  author  = {Kurt Hornik},
  title   = {Approximation capabilities of multilayer feedforward networks},
  journal = {Neural Networks},
  volume  = {4},
  number  = {2},
  pages   = {251--257},
  year    = {1991},
  doi     = {10.1016/0893-6080(91)90009-T},
  url     = {https://www.sciencedirect.com/science/article/pii/089360809190009T}
}

@inproceedings{shallowdeep,
  author    = {Delalleau, Olivier and Bengio, Yoshua},
  title     = {Shallow vs. Deep Sum-Product Networks},
  booktitle = {Proceedings of the 25th International Conference on Neural Information Processing Systems},
  series    = {NIPS'11},
  pages     = {666--674},
  numpages  = {9},
  publisher = {Curran Associates Inc.},
  address   = {Red Hook, NY, USA},
  location  = {Granada, Spain},
  year      = {2011}
}

@inproceedings{branchy,
  author    = {Teerapittayanon, Surat and McDanel, Bradley and Kung, H. T.},
  title     = {{BranchyNet}: Fast Inference via Early Exiting from Deep Neural Networks},
  booktitle = {2016 23rd International Conference on Pattern Recognition (ICPR)},
  pages     = {2464--2469},
  year      = {2016},
  doi       = {10.1109/ICPR.2016.7900006}
}

@inproceedings{ATHEENA,
  author    = {Biggs, Benjamin and Bouganis, Christos-Savvas and Constantinides, George A.},
  title     = {{ATHEENA}: A Toolflow for Hardware Early-Exit Network Automation},
  booktitle = {2023 IEEE 31st Annual International Symposium on Field-Programmable Custom Computing Machines (FCCM)},
  pages     = {121--132},
  year      = {2023},
  month     = {May},
  doi       = {10.1109/FCCM57271.2023.00022}
}

@inproceedings{fpgaconvnet,
  author    = {Venieris, Stylianos I. and Bouganis, Christos-Savvas},
  title     = {{fpgaConvNet}: Automated Mapping of Convolutional Neural Networks on {FPGAs} (Abstract Only)},
  booktitle = {Proceedings of the 2017 ACM/SIGDA International Symposium on Field-Programmable Gate Arrays},
  series    = {FPGA '17},
  pages     = {291--292},
  numpages  = {2},
  publisher = {Association for Computing Machinery},
  address   = {New York, NY, USA},
  location  = {Monterey, California, USA},
  year      = {2017},
  doi       = {10.1145/3020078.3021791},
  url       = {https://doi.org/10.1145/3020078.3021791}
}

@inproceedings{finn,
  author    = {Umuroglu, Yaman and others},
  title     = {{FINN}: A Framework for Fast, Scalable Binarized Neural Network Inference},
  booktitle = {Proceedings of the 2017 ACM/SIGDA International Symposium on Field-Programmable Gate Arrays},
  series    = {FPGA '17},
  pages     = {65--74},
  publisher = {Association for Computing Machinery},
  address   = {New York, NY, USA},
  location  = {Monterey, CA, USA},
  year      = {2017},
  month     = {Feb.},
  doi       = {10.1145/3020078.3021744}
}

@article{hls4ml,
  author        = {Duarte, Javier and others},
  title         = {Fast Inference of Deep Neural Networks in {FPGAs} for Particle Physics},
  journal       = {JINST},
  volume        = {13},
  number        = {07},
  pages         = {P07027},
  year          = {2018},
  doi           = {10.1088/1748-0221/13/07/P07027},
  eprint        = {1804.06913},
  archivePrefix = {arXiv},
  primaryClass  = {physics.ins-det}
}

@article{hls4ml-bin,
  author        = {Ngadiuba, Jennifer and others},
  title         = {Compressing Deep Neural Networks on {FPGAs} to Binary and Ternary Precision with {HLS4ML}},
  journal       = {Mach. Learn. Sci. Tech.},
  volume        = {2},
  pages         = {015001},
  year          = {2021},
  doi           = {10.1088/2632-2153/aba042},
  eprint        = {2003.06308},
  archivePrefix = {arXiv},
  primaryClass  = {cs.LG}
}

@inproceedings{hpipe,
  author    = {Hall, Mathew and Betz, Vaughn},
  title     = {{HPIPE}: Heterogeneous Layer-Pipelined and Sparse-Aware {CNN} Inference for {FPGAs}},
  booktitle = {Proceedings of the 2020 ACM/SIGDA International Symposium on Field-Programmable Gate Arrays},
  series    = {FPGA '20},
  pages     = {320},
  numpages  = {1},
  publisher = {Association for Computing Machinery},
  address   = {New York, NY, USA},
  location  = {Seaside, CA, USA},
  year      = {2020},
  doi       = {10.1145/3373087.3375380},
  url       = {https://doi.org/10.1145/3373087.3375380}
}

@inproceedings{h2pipe,
  author    = {Doumet, Mario and Stan, Marius and Hall, Mathew and Betz, Vaughn},
  title     = {{H2PIPE}: High Throughput {CNN} Inference on {FPGAs} with High-Bandwidth Memory},
  booktitle = {2024 34th International Conference on Field-Programmable Logic and Applications (FPL)},
  pages     = {69--77},
  publisher = {IEEE Computer Society},
  address   = {Los Alamitos, CA, USA},
  year      = {2024},
  month     = {Sep.},
  doi       = {10.1109/FPL64840.2024.00019},
  url       = {https://doi.ieeecomputersociety.org/10.1109/FPL64840.2024.00019}
}

@inproceedings{logicnets,
  author    = {Umuroglu, Yaman and Akhauri, Yash and Fraser, Nicholas J. and Blott, Michaela},
  title     = {{LogicNets}: Co-Designed Neural Networks and Circuits for Extreme-Throughput Applications},
  booktitle = {2020 30th International Conference on Field-Programmable Logic and Applications (FPL)},
  pages     = {291--297},
  publisher = {IEEE Computer Society},
  address   = {Gothenburg, Sweden},
  year      = {2020},
  month     = {Sep.},
  doi       = {10.1109/FPL50879.2020.00055}
}

@inproceedings{poly,
  author    = {Andronic, Marta and Constantinides, George A.},
  title     = {{PolyLUT}: Learning Piecewise Polynomials for Ultra-Low Latency {FPGA} {LUT}-based Inference},
  booktitle = {2023 International Conference on Field Programmable Technology (ICFPT)},
  pages     = {60--68},
  publisher = {IEEE},
  address   = {Yokohama, Japan},
  year      = {2023},
  month     = {Dec.},
  doi       = {10.1109/ICFPT59805.2023.00012}
}

@article{assemble2,
  author  = {Andronic, Marta and Li, Jiawen and Constantinides, George A.},
  title   = {{PolyLUT}: Ultra-Low Latency Polynomial Inference with Hardware-Aware Structured Pruning},
  journal = {{IEEE} Transactions on Computers},
  volume  = {74},
  number  = {9},
  pages   = {3181--3194},
  year    = {2025},
  doi     = {10.1109/TC.2025.3586311}
}

@inproceedings{NeuraLUT,
  author    = {Andronic, Marta and Constantinides, George A.},
  title     = {{NeuraLUT}: Hiding Neural Network Density in Boolean Synthesizable Functions},
  booktitle = {2024 34th International Conference on Field-Programmable Logic and Applications (FPL)},
  pages     = {140--148},
  publisher = {IEEE},
  year      = {2024},
  doi       = {10.1109/FPL64840.2024.00028}
}

@inproceedings{assemble,
  author    = {Andronic, Marta and Constantinides, George A.},
  title     = {{NeuraLUT-Assemble}: Hardware-Aware Assembling of Sub-Neural Networks for Efficient {LUT} Inference},
  booktitle = {2025 IEEE 33rd Annual International Symposium on Field-Programmable Custom Computing Machines (FCCM)},
  pages     = {208--216},
  publisher = {IEEE},
  year      = {2025},
  doi       = {10.1109/FCCM62733.2025.00077}
}

@inproceedings{treelut,
  author    = {Khataei, Alireza and Bazargan, Kia},
  title     = {{TreeLUT}: An Efficient Alternative to Deep Neural Networks for Inference Acceleration Using Gradient Boosted Decision Trees},
  booktitle = {Proceedings of the 2025 ACM/SIGDA International Symposium on Field Programmable Gate Arrays},
  series    = {FPGA '25},
  pages     = {14--24},
  numpages  = {11},
  publisher = {Association for Computing Machinery},
  address   = {New York, NY, USA},
  location  = {Monterey, CA, USA},
  year      = {2025},
  doi       = {10.1145/3706628.3708877},
  url       = {https://doi.org/10.1145/3706628.3708877}
}

@inproceedings{amigolut,
  author    = {Weng, Olivia and Andronic, Marta and Zuberi, Danial and Chen, Jiaqing and Geniesse, Caleb and Constantinides, George A. and Tran, Nhan and Fraser, Nicholas J. and Duarte, Javier Mauricio and Kastner, Ryan},
  title     = {Greater than the Sum of Its {LUTs}: Scaling Up {LUT}-based Neural Networks with {AmigoLUT}},
  booktitle = {Proceedings of the 2025 ACM/SIGDA International Symposium on Field Programmable Gate Arrays},
  series    = {FPGA '25},
  pages     = {25--35},
  numpages  = {11},
  publisher = {Association for Computing Machinery},
  address   = {New York, NY, USA},
  location  = {Monterey, CA, USA},
  year      = {2025},
  doi       = {10.1145/3706628.3708874},
  url       = {https://doi.org/10.1145/3706628.3708874}
}

@inproceedings{dwn,
  author    = {Bacellar, Alan Tendler Leibel and Susskind, Zachary and Breternitz Jr, Mauricio and John, Eugene and John, Lizy Kurian and Lima, Priscila Machado Vieira and Fran\c{c}a, Felipe M. G.},
  title     = {Differentiable Weightless Neural Networks},
  booktitle = {Proceedings of the 41st International Conference on Machine Learning},
  series    = {Proceedings of Machine Learning Research},
  volume    = {235},
  pages     = {2277--2295},
  publisher = {PMLR},
  year      = {2024},
  month     = {21--27 Jul},
  url       = {https://proceedings.mlr.press/v235/bacellar24a.html}
}

@inproceedings{kanele,
  author    = {Hoang, Duc and Gupta, Aarush and Harris, Philip C.},
  title     = {{KANEL{\'E}}: {Kolmogorov--Arnold} Networks for Efficient {LUT}-based Evaluation},
  booktitle = {Proceedings of the 2026 ACM/SIGDA International Symposium on Field Programmable Gate Arrays},
  pages     = {44--55},
  publisher = {Association for Computing Machinery},
  year      = {2026},
  month     = {Feb.},
  doi       = {10.1145/3748173.3779202},
  url       = {https://doi.org/10.1145/3748173.3779202}
}

@misc{hgqlut,
  author        = {Sun, Chang and Que, Zhiqiang and Zadeh, Bakhtiar and Liu, Qibin and Alvarez, Kevin H. and Luk, Wayne and Spiropulu, Maria},
  title         = {{HGQ-LUT}: Fast {LUT}-Aware Training and Efficient Architectures for {DNN} Inference},
  year          = {2026},
  eprint        = {2604.22293},
  archivePrefix = {arXiv},
  primaryClass  = {cs.AR},
  url           = {https://arxiv.org/abs/2604.22293}
}

@inproceedings{yukio,
  author    = {Miyasaka, Y. and Mishchenko, A. and Wawrzynek, J. and Fraser, N. J.},
  title     = {Synthesis of {LUT} Networks for Random-Looking Dense Functions with Don't Cares: Towards Efficient {FPGA} Implementation of {DNN}},
  booktitle = {2024 IEEE 32nd Annual International Symposium on Field-Programmable Custom Computing Machines (FCCM)},
  pages     = {126--132},
  publisher = {IEEE Computer Society},
  address   = {Los Alamitos, CA, USA},
  year      = {2024},
  month     = {May},
  doi       = {10.1109/FCCM60383.2024.00023}
}

@inproceedings{compressedlut,
  author    = {Khataei, Alireza and Bazargan, Kia},
  title     = {{CompressedLUT}: An Open Source Tool for Lossless Compression of Lookup Tables for Function Evaluation and Beyond},
  booktitle = {Proceedings of the 2024 ACM/SIGDA International Symposium on Field Programmable Gate Arrays},
  series    = {FPGA '24},
  pages     = {2--11},
  numpages  = {10},
  publisher = {Association for Computing Machinery},
  address   = {New York, NY, USA},
  location  = {Monterey, CA, USA},
  year      = {2024},
  doi       = {10.1145/3626202.3637575},
  url       = {https://doi.org/10.1145/3626202.3637575}
}

@inproceedings{reducedlut,
  author    = {Cassidy, Oliver and Andronic, Marta and Coward, Samuel and Constantinides, George A.},
  title     = {{ReducedLUT}: Table Decomposition with ``Don't Care'' Conditions},
  booktitle = {Proceedings of the 2025 ACM/SIGDA International Symposium on Field Programmable Gate Arrays},
  series    = {FPGA '25},
  pages     = {36--42},
  numpages  = {7},
  publisher = {Association for Computing Machinery},
  address   = {New York, NY, USA},
  location  = {Monterey, CA, USA},
  year      = {2025},
  doi       = {10.1145/3706628.3708823},
  url       = {https://doi.org/10.1145/3706628.3708823}
}

@inproceedings{sdn,
  author    = {Kaya, Yigitcan and Hong, Sanghyun and Dumitras, Tudor},
  title     = {Shallow-Deep Networks: Understanding and Mitigating Network Overthinking},
  booktitle = {Proceedings of the 36th International Conference on Machine Learning},
  series    = {Proceedings of Machine Learning Research},
  volume    = {97},
  pages     = {3301--3310},
  publisher = {PMLR},
  year      = {2019},
  month     = {09--15 Jun},
  url       = {https://proceedings.mlr.press/v97/kaya19a.html}
}

@inproceedings{msd,
  author    = {Huang, Gao and Chen, Danlu and Li, Tianhong and Wu, Felix and van der Maaten, Laurens and Weinberger, Kilian},
  title     = {Multi-Scale Dense Networks for Resource Efficient Image Classification},
  booktitle = {International Conference on Learning Representations},
  year      = {2018},
  url       = {https://openreview.net/forum?id=Hk2aImxAb}
}

@inproceedings{sgd,
  author    = {Loshchilov, Ilya and Hutter, Frank},
  title     = {{SGDR}: Stochastic Gradient Descent with Warm Restarts},
  booktitle = {5th International Conference on Learning Representations},
  address   = {Toulon, France},
  year      = {2017},
  month     = {Apr.},
  url       = {https://openreview.net/forum?id=Skq89Scxx}
}

@inproceedings{decoupled,
  author    = {Loshchilov, Ilya and Hutter, Frank},
  title     = {Decoupled Weight Decay Regularization},
  booktitle = {International Conference on Learning Representations},
  year      = {2019},
  url       = {https://openreview.net/forum?id=Bkg6RiCqY7}
}

@misc{brevitas,
  author = {Pappalardo, Alessandro},
  title  = {{Brevitas}: Neural Network Quantization in {PyTorch}},
  year   = {2020},
  month  = {Jun.},
  note   = {[Accessed: November 5, 2025]},
  url    = {https://github.com/Xilinx/brevitas}
}

@inproceedings{deep-nets,
  author    = {Lee, Chen-Yu and Xie, Saining and Gallagher, Patrick and Zhang, Zhengyou and Tu, Zhuowen},
  title     = {Deeply-Supervised Nets},
  booktitle = {Proceedings of the Eighteenth International Conference on Artificial Intelligence and Statistics},
  series    = {Proceedings of Machine Learning Research},
  volume    = {38},
  pages     = {562--570},
  publisher = {PMLR},
  address   = {San Diego, California, USA},
  year      = {2015},
  month     = {09--12 May},
  url       = {https://proceedings.mlr.press/v38/lee15a.html}
}

@inproceedings{triplewins,
  author    = {Hu, Ting-Kuei and Chen, Tianlong and Wang, Haotao and Wang, Zhangyang},
  title     = {Triple Wins: Boosting Accuracy, Robustness and Efficiency Together by Enabling Input-Adaptive Inference},
  booktitle = {International Conference on Learning Representations},
  year      = {2020},
  url       = {https://openreview.net/forum?id=rJgzzJHtDB}
}

@misc{mnist,
  author = {LeCun, Yann and Cortes, Corinna and Burges, Christopher J. C.},
  title  = {The {MNIST} Database of Handwritten Digits},
  url    = {http://yann.lecun.com/exdb/mnist/}
}

@misc{cifar10,
  author       = {Krizhevsky, Alex and Nair, Vinod and Hinton, Geoffrey},
  title        = {Learning Multiple Layers of Features from Tiny Images},
  year         = {2009},
  howpublished = {\url{https://www.cs.toronto.edu/~kriz/cifar.html}},
  note         = {[Accessed: May 17, 2026]}
}

@misc{cernbox_jsc,
  author       = {{CERN Collaboration}},
  title        = {{CERNBox} {LHC} Jets Dataset},
  year         = {2025},
  howpublished = {\url{https://cernbox.cern.ch/index.php/s/jvFd5MoWhGs1l5v/download}},
  note         = {[Accessed: November 6, 2025]}
}

@misc{openml_jsc,
  author       = {{OpenML Contributors and LHC Jets HLF Curators}},
  title        = {{hls4ml\_lhc\_jets\_hlf} ({OpenML} Dataset 42468)},
  year         = {2020},
  howpublished = {\url{https://www.openml.org/d/42468}},
  note         = {[Accessed: November 11, 2025]}
}

\end{document}